\documentclass[]{IEEEtran}
\usepackage{cite}
\usepackage{graphicx}
\usepackage{amsmath}
\usepackage{times}
\usepackage{latexsym}
\usepackage{graphicx}
\usepackage{bm}
\usepackage{amssymb}
\usepackage{stfloats}
\usepackage{cases}
\usepackage{array}
\usepackage{setspace}
\usepackage{fancyhdr}
\usepackage{citesort}
\usepackage{color}
\usepackage{diagbox}
\usepackage[center]{caption}
\usepackage[margin=0.7in,top=0.87in]{geometry}

\usepackage{algorithm}
\usepackage{algorithmicx}
\usepackage{algpseudocode}
\usepackage{graphicx} 
\usepackage{epstopdf}
\usepackage{arydshln}
\allowdisplaybreaks[4]

\usepackage{graphicx}
\usepackage{subfig}

\begin{document}
	
	\captionsetup{font={small}}
	\vspace{-1ex}
\title{\huge Meta-Learning-Driven Adaptive Codebook Design for Near-Field Communications}
\author{Mianyi Zhang$^1$, Yunlong Cai$^1$, Jiaqi Xu$^2$, and A. Lee Swindlehurst$^2$
	\\$^1$ Zhejiang University, Hangzhou 310027, China
	\\$^2$University of California at Irvine, Irvine, CA 92697, USA
	\\mianyi\_zhang@zju.edu.cn, ylcai@zju.edu.cn, xu.jiaqi@uci.edu, swindle@uci.edu
\thanks{}
\vspace{-2ex}
}

\maketitle
\thispagestyle{empty}
\begin{abstract}
	Extremely large-scale arrays (XL-arrays) and ultra-high frequencies are two key technologies for sixth-generation (6G) networks, offering higher system capacity and expanded bandwidth resources. To effectively combine these technologies, it is necessary to consider the near-field spherical-wave propagation model, rather than the traditional far-field planar-wave model. In this paper, we explore a near-field communication system comprising a base station (BS) with hybrid analog-digital beamforming and multiple mobile users. Our goal is to maximize the system's sum-rate by optimizing the near-field codebook design for hybrid precoding. To enable fast adaptation to varying user distributions, we propose a meta-learning-based framework that integrates the model-agnostic meta-learning (MAML) algorithm with a codebook learning network. Specifically, we first design a deep neural network (DNN) to learn the near-field codebook. Then, we combine the MAML algorithm with the DNN to allow rapid adaptation to different channel conditions by leveraging a well-initialized model from the outer network. Simulation results demonstrate that our proposed framework outperforms conventional algorithms, offering improved generalization and better overall performance.


\end{abstract}\label{key}

\section{Introduction}

The growing demand for higher data transmission rates and enhanced spatial resolution in wireless communication systems has accelerated the development of extremely large-scale arrays (XL-arrays) and the adoption of very high frequencies \cite{NF1}. XL-arrays are capable to provide remarkable system capacity and spatial resolution \cite{XL1,XL2,XL3}. Additionally, operating at higher frequencies offers expanded bandwidth and facilitates the deployment of XL-arrays due to the shorter wavelengths \cite{THz1,THz2}. However, integrating these technologies requires a significant shift from conventional far-field to near-field communication.

\begin{figure}[h]
	\vspace{0ex}
	\centering
	\includegraphics[scale=0.25]{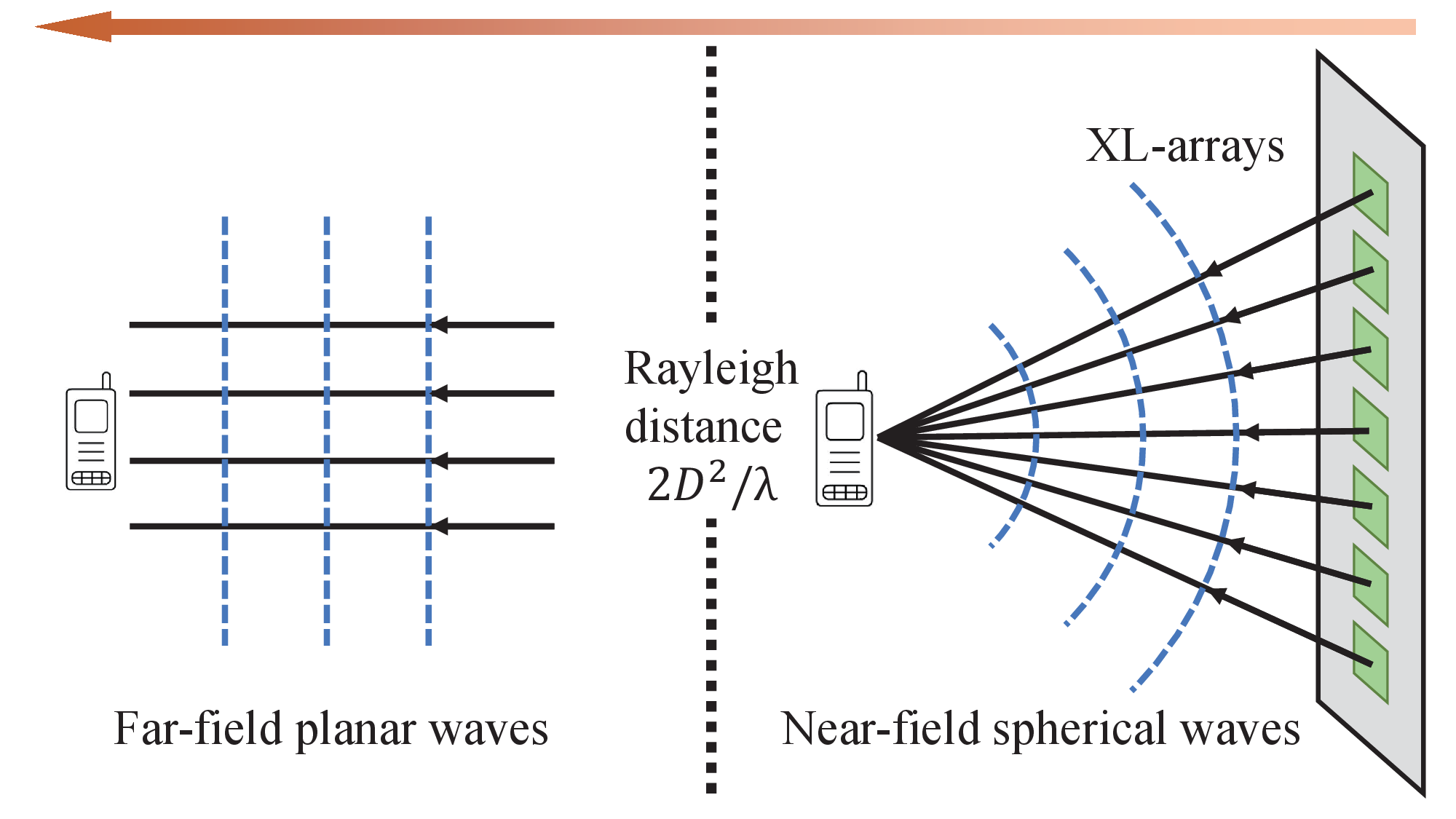}
	\vspace{2ex}
	\caption{The near-field region and the far-field region divided by the Rayleigh distance.}
	\label{nearfield}
	\vspace{0ex}
\end{figure}

As illustrated in Fig. \ref{nearfield}, unlike far-field communication, where signals propagate as planar waves, near-field communication is characterized by spherical wave propagation \cite{NF2,NF3}. The boundary between the two regions is typically defined by the Rayleigh distance, calculated as $2D^2/\lambda$, where $D$ and $\lambda$ denote the antenna aperture and the wavelength, respectively \cite{NF3}. In current 5G wireless networks, communication systems are generally designed based on far-field approximations, as the near-field effects are minimal due to the smaller antenna arrays and lower frequencies. However, with the larger apertures of XL-arrays and higher frequencies anticipated in future 6G networks, the near-field region can extend beyond 100 meters, which indicates that the conventional far-field plane wave assumption becomes unsuitable, and near-field effects must be taken into account \cite{NF4,NF5}.

Several studies have explored near-field communications, addressing areas such as near-field beamfocusing \cite{NFBF}, channel estimation \cite{NFCE}, integrated sensing and communication \cite{NFISAC}, and reconfigurable intelligent surfaces \cite{NFRIS}. Additionally, hybrid analog-digital beamforming is a crucial technology in near-field communication, necessitated by the use of XL-arrays and extremely high carrier frequencies \cite{NF1}. In \cite{HB1}, the authors designed a hierarchical codebook for near-field communications, proposing an approximation method for the steering beam gain in lower-layer codebook, and methods for beam rotation and relocation in the upper-layer codebook. Meanwhile, the authors in \cite{HB2} introduced a two-phase beam training design. In the first phase, the candidate angle is obtained based on traditional beam training for far-field scenarios. Then, the distance in polar-domain is estimated in the second phase.

However, to the best of authors' knowledge, learnable codebook design has not yet been explored in near-field communications. Traditional beamforming codebooks such as those based on the Discrete Fourier Transform (DFT) \cite{DFT}, are designed to work in far-field regimes by generating massive beams covering the entire beam domain. These codebooks have several drawbacks: (i) they work in the angle domain and are not suitable in near-field scenarios where the beamforming is affected by both the angle and distance, referred to as beam-focusing, (ii) even with near-field beam-focusing, traditional codebooks uniformly designed on both the angle and distance may not be able to provide an acceptable beam density at the users' locations, (iii) they waste beam resources due to the nonuniform distribution of users, since beams steered towards areas distant from the users may never be used, and (iv) they are typically predefined and struggle to adapt to environmental changes and user dynamics. To overcome these limitations, machine learning-based codebook design presents a promising solution \cite{MLCB0}. For instance, in \cite{MLCB}, the authors proposed a machine learning-enabled framework for far-field codebook design and demonstrated the advantages of a learnable codebook.

In this paper, we examine a near-field communication system consisting of a base station (BS) equipped with a large-scale uniform planar array (UPA) and multiple mobile users. The BS employs hybrid analog-digital beamforming, and we aim to optimize the system's sum-rate by designing an efficient near-field codebook. To account for environmental changes and user dynamics, we propose a meta-learning framework that integrates a model-agnostic meta-learning (MAML) algorithm with a codebook learning network. Specifically, we first develop a deep neural network (DNN) to learn the near-field codebook. The MAML algorithm is then carefully integrated with this DNN, where the DNN serves as the inner network. In this framework, the outer network enables the inner network to achieve effective initialization, allowing for rapid adaptation to different channels. Simulation results show that our approach outperforms conventional algorithms, offering superior generalization.

\begin{figure*}[ht]
	\vspace{3ex}
	\centering
	\includegraphics[scale=0.53]{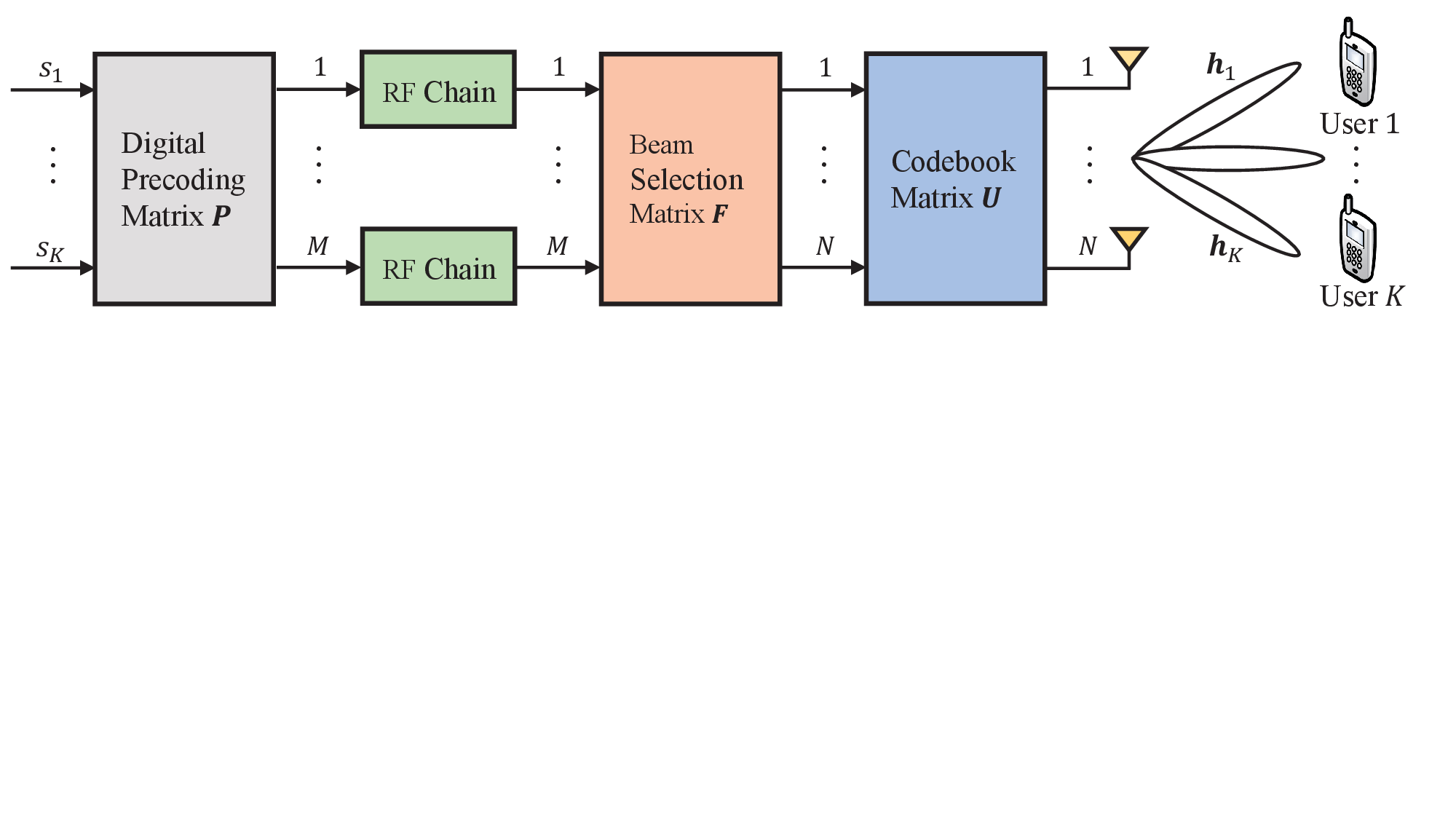}
	\vspace{-37ex}
	\caption{Architectures for hybrid analog-digital beamforming.}
	\label{system}
	\vspace{0ex}
\end{figure*}

{\it Notation}: The symbols ${\left\| {\cdot} \right\|}$, $({\cdot})^{T}$, and $({\cdot})^{H}$ denote the Euclidean norm, transpose operator, complex conjugate transpose operator, respectively. ${\rm{Tr}}(\cdot)$ stands for the trace of a matrix. Besides, the spaces of $x\times y$ real- and complex-values matrices are denoted by ${\mathbb R^{x \times y}}$ and ${\mathbb C^{x \times y}}$ respectively, and $\bm{I}$ represents the identity matrix. The complex Gaussian distribution with zero mean and covariance matrix $\bm{A}$ is denoted by $\mathcal{CN}(\bm{0},\bm{A})$. Additionally, the expectation of the random variable $x$ is indicated by $\mathbb{E}\{x\}$, and $[x]_+$ represents the nonnegative part of $x$.

\begin{figure}[h]
	\vspace{0ex}
	\centering
	\includegraphics[scale=0.3]{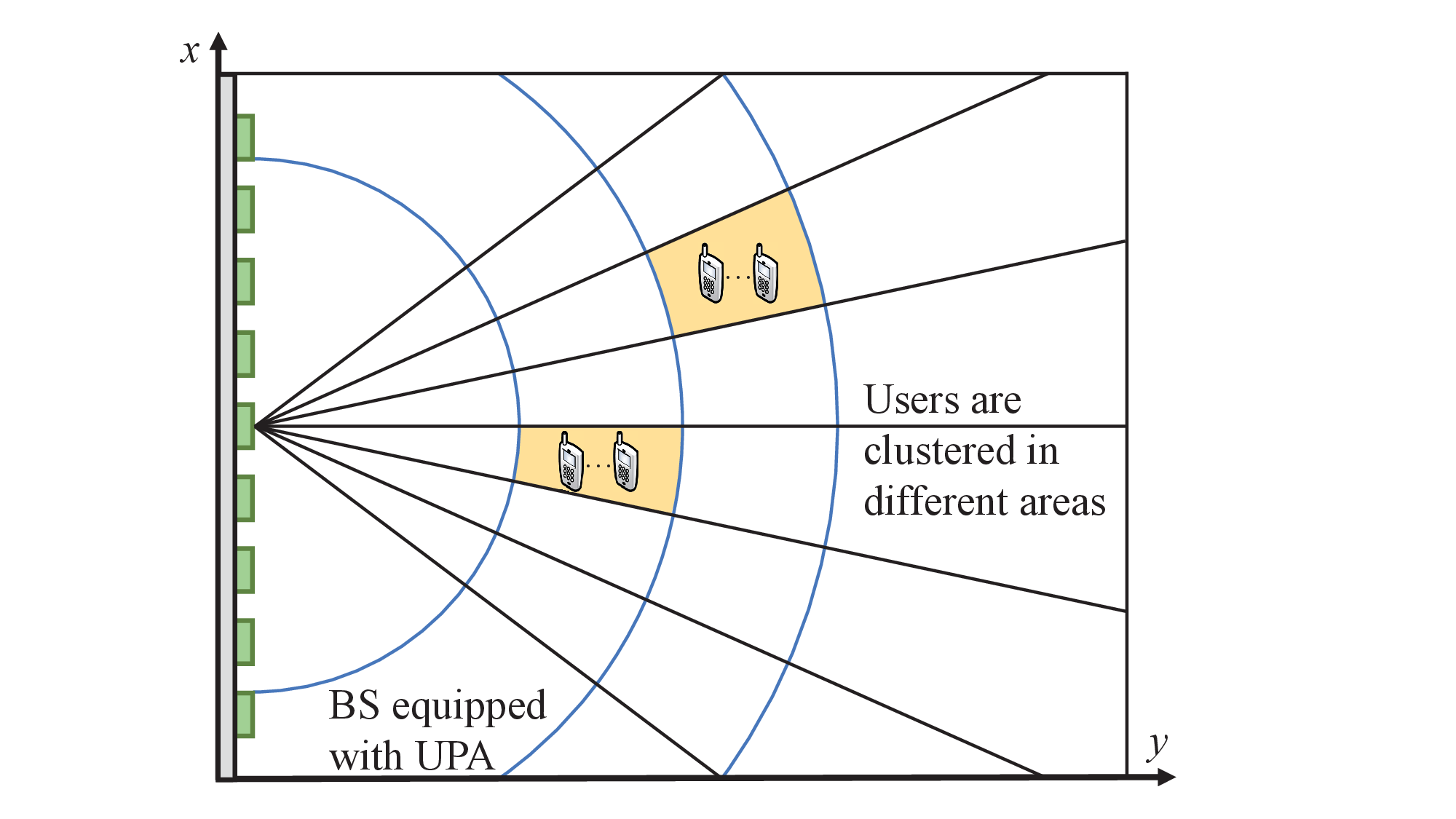}
	\vspace{-2ex}
	\caption{A near-field multi-user communication system where users are clustered in different areas.}
	\label{polar}
	\vspace{0ex}
\end{figure}

\section{System Model and Problem Formulation}

In this section, we begin by presenting the near-field communication system, followed by the formulation of the optimization problem addressed in this paper.

\subsection{System Model}

As shown in Fig. \ref{polar}, we consider a near-field multi-user system comprising a BS equipped with a UPA consisting of $N=N_xN_z$ transmit antennas, serving $K$ single-antenna mobile users distributed across different areas. Additionally, the BS employs hybrid analog-digital beamforming with $M \ll N$ RF-chains, as depicted in Fig. \ref{system}. The precoded data vector at the BS is then expressed as
\begin{align}
	{\bm x}={\bm P}{\bm s}=\sum\limits^{K}_{k=1}{\bm p}_ks_k,
\end{align}
where ${\bm s} \triangleq [s_1,s_2,...,s_K]^T$, $s_k$ denotes the transmit signal for user $k$ with zero mean and $\mathbb{E}[{\bm s}{\bm s}^H]={\bm I}$, ${\bm P}\triangleq [{\bm p}_1,{\bm p}_2,...,{\bm p}_K]\in {\mathbb C}^{M\times K}$ to denote the digital precoder, and ${\bm p}_k$ stands for the digital precoding vector for user $k$. Then, the received signal for user $k$ is given by
\begin{align}
	y_k=
	{\bm h}_{k}^H{\bm U}{\bm F}{\bm P}{\bm s}+n_k,
\end{align}
where ${\bm h}_{k}$ denotes the channel vector from the BS to user $k$, ${\bm U} \triangleq [{\bm u}_1,...,{\bm u}_N]\in {\mathbb C}^{N\times N} $ denotes the codebook matrix, and $\bm F$ denotes the beam selection matrix whose entries $f_{ij}, (i,j) \in \mathcal T$ are either $0$ or $1$, and $\mathcal T \triangleq \{(i,j)|i=1,...,N, j=1,...,M\}$. Let $N_x=2\tilde{N}_x+1$ and $N_z=2\tilde{N}_z+1$. Then ${\bm u}_n = \dfrac{1}{\sqrt{N}} [e^{j\theta^n_{-\tilde{N}_x,-\tilde{N}_z}},...,e^{j\theta^n_{\tilde{N}_x,\tilde{N}_z}}]^T$ denotes the $n$-th beam vector in the codebook, and $\theta^n_{n_x,n_z} \in \left[ 0,2\pi \right)$ denotes the phase shift of the corresponding antenna. Finally, $n_k \sim \mathcal{CN}(\bm{0},\sigma^2)$ represents additive white Gaussian noise.

Let $L$ denote the total number of scatterers, ${\bm s}_{n_x,n_z}$ denotes the coordinates of the antenna at position $n_x, n_z$, ${\bm r}_k$ and ${\bm r}_l$ denotes the coordinates of user $k$ and the $l$-th scatterer, respectively. Additionally, let $\beta_k$ and $\beta_l$ denote the channel gains of user $k$ and the $l$-th scatterer, and let $\tilde{\beta}_l$ represent the i.i.d. random phase of the $l$-th scatterer. The channel vectors between the BS and the users are modelled as follows:
\begin{align}
	{\bm h}_{k}=\beta_k {\bm a}({\bm r}_k) + \sum\limits^{L}_{l=1} \tilde{\beta}_l\beta_l {\bm a}({\bm r}_l),
\end{align}
where ${\bm a}({\bm r}_k) \triangleq [e^{-j\frac{2\pi}{\lambda}(\|{\bm r}_k-{\bm s}_{-\tilde{N}_x,-\tilde{N}_z}\|-r_k)},...,\\\notag e^{-j\frac{2\pi}{\lambda}(\|{\bm r}_k-{\bm s}_{\tilde{N}_x,\tilde{N}_z}\|-r_k)}]^T$ denotes the BS array response vector, $r_k \triangleq \|{\bm r}_k-{\bm s}_{0,0}\|$, and $\beta_k \triangleq \dfrac{1}{\sqrt{4\pi r_k^2}}e^{-j\frac{2\pi}{\lambda}r_k}$.

\begin{figure*}[t]
	\vspace{0ex}
	\centering
	\includegraphics[scale=0.53]{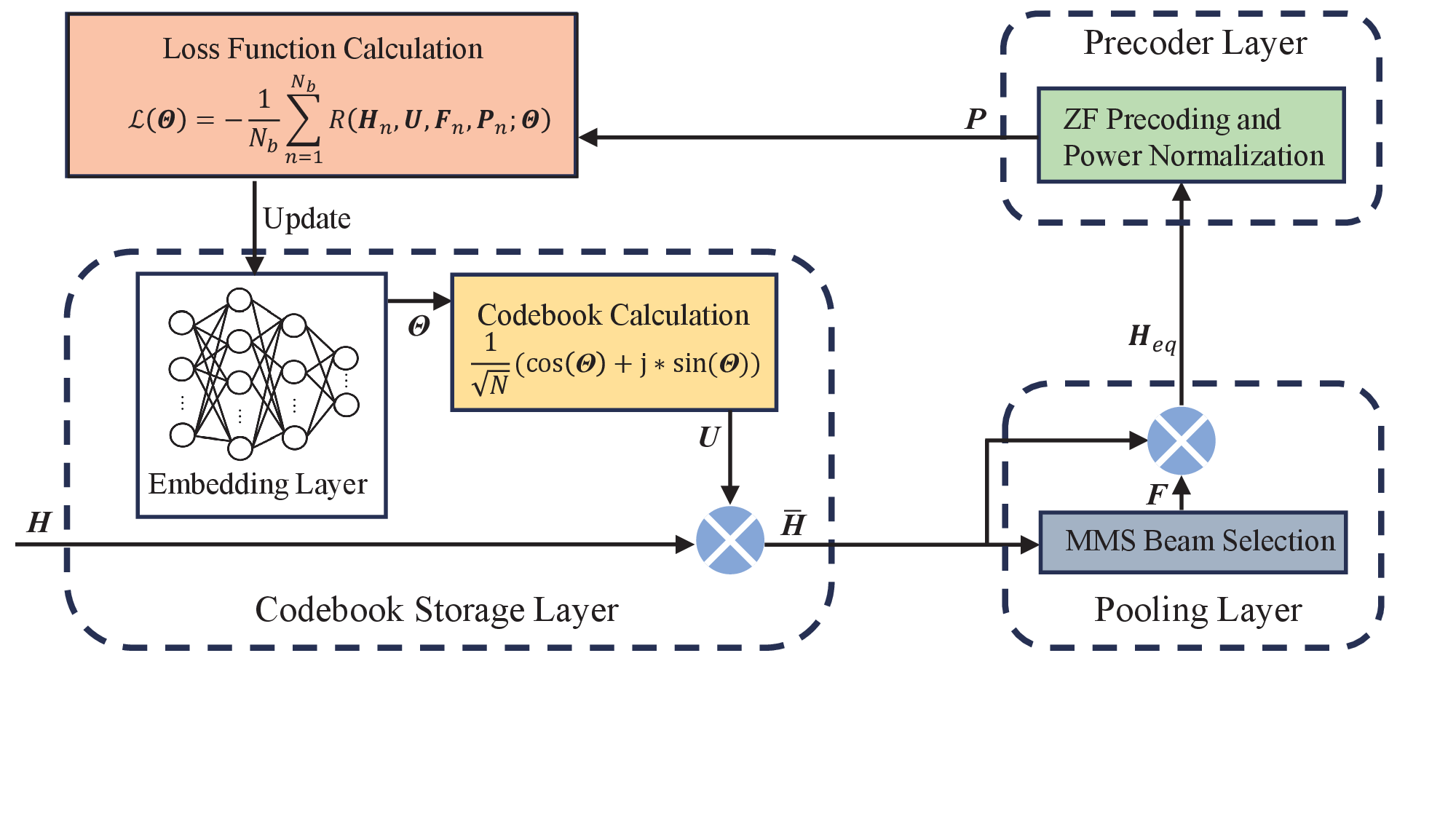}
	\vspace{-12ex}
	\caption{Architectures of codebook design DNN.}
	\label{DNN}
	\vspace{0ex}
\end{figure*}

\subsection{Problem Formulation}
In this paper, we focus on maximizing the sum-rate of the system by optimizing the near-field codebook matrix ${\bm U}$, the beam selection matrix ${\bm F}$, and the digital precoding vector ${\bm p}_k$. We formulate the system sum-rate maximization problem as

\begin{subequations} \label{p1}
	\begin{eqnarray}
		&\max\limits_{\{{\bm U},{\bm F},{\bm p}_k\}}& \sum\limits^{K}_{k=1}\log_2\left(1\!+\!\frac{|{\bm h}_{k}^H{\bm U}{\bm F}{\bm p}_k|^2}{{\sum\limits^{K}_{i \ne k}
				|{\bm h}_{k}^H{\bm U}{\bm F}{\bm p}_i|^2}\!+\!\sigma^2}\right)\\
		&\text{s.t.}&\sum\limits^{K}_{k=1}\|{\bm p}_k\|^2 \leq P_{\text{max}},\\
		&&\sum\limits^{N}_{i=1}f_{ij}=1,\forall j,\\
		&&\sum\limits^{M}_{j=1}f_{ij} \leq 1,\forall i,\\
		&&f_{ij} \in \{0,1\}, \forall (i,j) \in \mathcal T,
	\end{eqnarray}
\end{subequations}
where (\ref{p1}c) is the power constraint and $P_{\text{max}}$ is the transmit power budget at the BS. Constraint (\ref{p1}c) ensures that each RF chain only selects one beam and constraint (\ref{p1}d) guarantees that each beam can only be selected by at most one RF chain. Using the designed codebook matrix ${\bm U}$, the beam selection matrix ${\bm F}$ is determined by applying the maximum magnitude selection (MMS) algorithm \cite{MMS}. The digital precoding vector ${\bm p}_k$ is then calculated using the zero-forcing (ZF) algorithm \cite{ZF}. To solve problem (\ref{p1}), we propose a meta-learning-based framework described in the following section.

\section{Proposed meta-learning based framework}

In this section, we propose a meta-learning based framework to solve problem (\ref{p1}).

\subsection{DNN for Learning Near-Field Codebook}

We begin by presenting the DNN architecture for learning the near-field codebook. As illustrated in Fig. \ref{DNN}, the network consists of three layers: the codebook storage layer, the pooling layer, and the precoder layer.

\begin{itemize}
	\item[$\bullet$] Codebook Storage Layer: Since the elements of the codebook matrix are all complex with unit magnitude, we first store the codebook phase shift matrix ${\bm \Theta}$ in the embedding layer as learnable parameters, where ${\bm \Theta}\triangleq [{\bm \theta}^1,...,{\bm \theta}^N] \in {\mathbb R}^{N\times N}$, ${\bm \theta}^n \triangleq [\theta^n_{-\tilde{N}_x,-\tilde{N}_z},...,\theta^n_{\tilde{N}_x,\tilde{N}_z}]^T$ denotes the beam phase vector, and $\theta^n_{n_x,n_z} \in \left[ 0,2\pi \right)$. Using Euler's formula and scaling by $1/\sqrt{N}$, the phase shift matrix ${\bm \Theta}$ can be transformed to the codebook matrix ${\bm U}$ as
	\begin{align}\label{U}
		{\bm U}=\frac{1}{\sqrt{N}}(\text{cos}({\bm \Theta})+j*\text{sin}({\bm \Theta})).
	\end{align}
Then, the beamspace channel matrix $\bar{\bm H}$ can be obtained by executing the matrix multiplication process within the layer as $\bar{\bm H}^H = {\bm H}^H{\bm U}$, where ${\bm H}\triangleq [{\bm h}_1,{\bm h}_2,...,{\bm h}_K]$.

	\item[$\bullet$] Pooling Layer: The beamspace channel matrix $\bar{\bm H}$ is the input to this layer, and the MMS algorithm is employed to obtain the beam selection matrix ${\bm F}$. The MMS algorithm evaluates the energy received by each user from the $N$ beams and selects the strongest $M$ beams. Then, the equivalent lower-dimensional channel matrix ${\bm H}_{eq}$ can be obtained as ${\bm H}_{eq}^H = \bar{\bm H}^H {\bm F}$.

	\item[$\bullet$] Precoder Layer: In this layer, the unnormalized digital precoding matrix $\bar{\bm P}$ is first obtained by applying zero-forcing (ZF) with the input equivalent channel matrix ${\bm H}_{eq}$:
		\begin{align}\label{zf1}
		\bar{\bm P}={\bm H}_{eq}({\bm H}_{eq}^H{\bm H}_{eq})^{-1}.
	\end{align}
The final digital precoder ${\bm P}$ is found by normalizing $\bar{\bm P}$:
\begin{align}\label{zf2}
{\bm P}=\frac{\sqrt{P_{\text{max}}}}{\sqrt{{\rm{Tr}}(\bar{\bm P}^H \bar{\bm P})}}\bar{\bm P}.
\end{align}
	
	\item[$\bullet$] Training Procedure and Loss Function: The training set $T_t$ includes channel matrices generated according to the channel model described in Section II. The model is trained through multiple iterations of forward and backward propagation. In each iteration, a batch of channel matrices randomly chosen from the training set is input into the model and processed through the above three layers. The loss function of the network is calculated as follows:
			\begin{align}
		\mathcal{L}({\bm \Theta})=-\frac{1}{N_b}\sum\limits^{N_b}_{n=1}R({\bm H}_n,{\bm U},{\bm F}_n,{\bm P}_n;{\bm \Theta}),
	\end{align}
where $N_b$ is the batch size, $R(\cdot)$ is the sum-rate function (\ref{p1}a), and ${\bm H}_n$ denotes the $n$-th sample in the batch. Given input ${\bm H}_n$, ${\bm F}_n$ and ${\bm P}_n$ can be obtained in the second and the third layer, respectively. Then, ${\bm \Theta}$ is updated by
			\begin{align}
{\bm \Theta}_{i+1}={\bm \Theta}_i-\alpha_1 \frac{\partial \mathcal{L}}{\partial{\bm \Theta}_i},
\end{align}
where $\alpha_1$ denotes the network learning rate. $\textbf{Algorithm 1}$ demonstrates the training procedure for the codebook design DNN.
\end{itemize}

\begin{algorithm}[h]
	\caption{Training Procedure for Codebook Design DNN}
	\begin{algorithmic}[1]
		\State {\bf Input}:\ batch size $N_b$, training set $T_t$, network learning rate $\alpha_1$.
		\State {\bf Initialize}:\ codebook phase shift matrix ${\bm \Theta}$.
		\State {\bf for} epoch $=1, 2, ...$ {\bf do}
		\State \quad Sample $N_b$ channel matrices;
		\State \quad Calculate the codebook matrix ${\bm U}$ based on (\ref{U});
		\State \quad Compute the beamspace channel matrix $\bar{\bm H}$;
		\State \quad Input $\bar{\bm H}$ into the pooling layer and apply the MMS 
		\Statex \quad \quad algorithm to obtain the beam selection matrix ${\bm F}$;
		\State \quad Calculate the equivalent channel matrix ${\bm H}_{eq}$;
		\State \quad Input ${\bm H}_{eq}$ into the precoder layer. The digital precoder 
		\Statex \quad \quad matrix ${\bm P}$ is obtained by employing ZF algorithm 
		\Statex \quad \quad as in (\ref{zf1}) and (\ref{zf2});
		\State \quad Compute $\mathcal{L}({\bm \Theta})$ and update ${\bm \Theta}$.
		\State {\bf end}
		\label{algorithm 1}
	\end{algorithmic}
\end{algorithm}

\begin{figure}[h]
	\vspace{-2ex}
	\centering
	\includegraphics[scale=0.24]{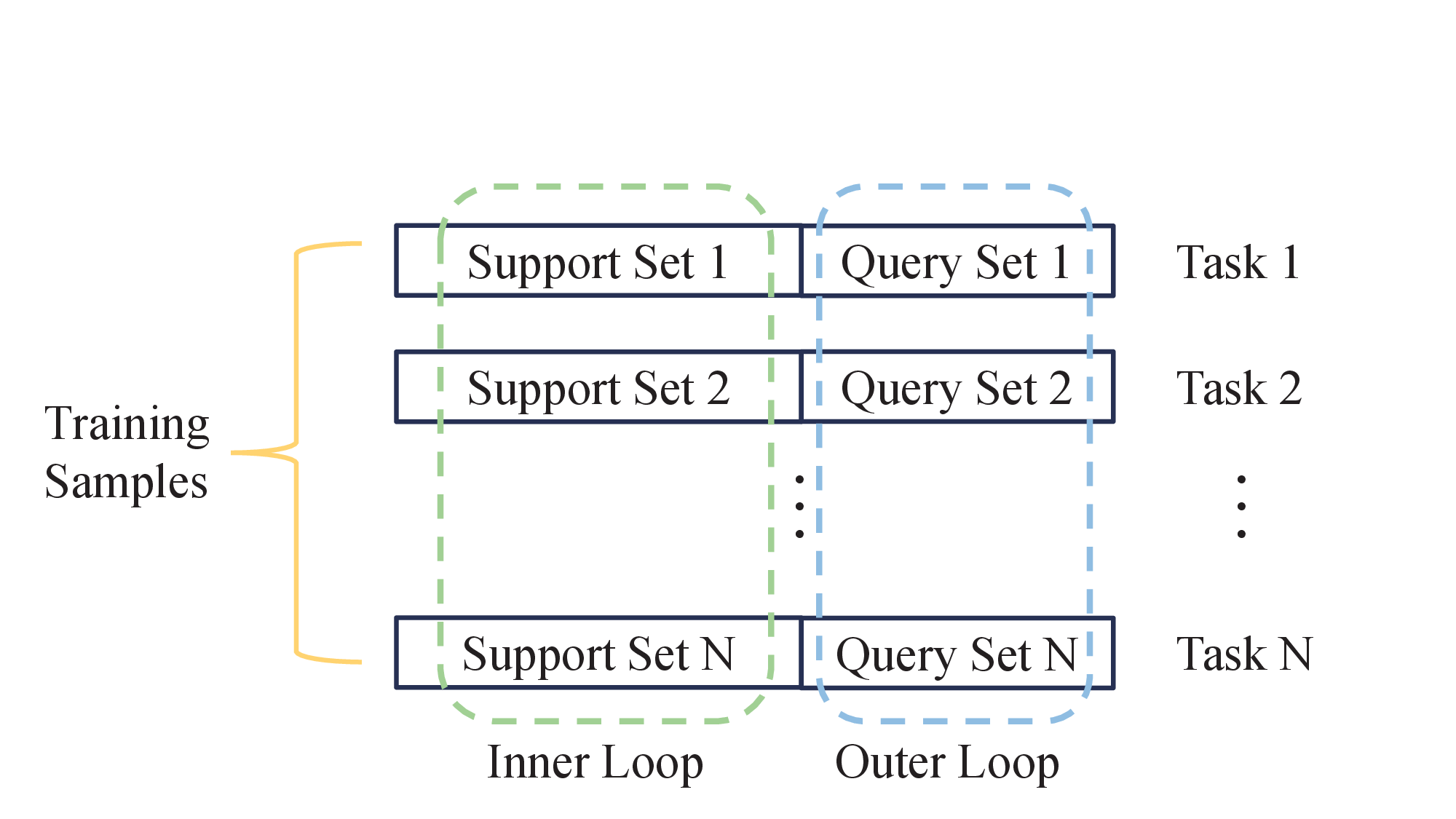}
	\vspace{-0ex}
	\caption{Construction of the dataset.}
	\label{MAML}
	\vspace{0ex}
\end{figure}

\subsection{MAML based Framework}

To enable the network to quickly adapt to varying user distributions, we propose a MAML-based framework. Unlike conventional deep learning methods where channel state information is directly used as training data, MAML treats training samples as tasks. As illustrated in Fig. \ref{MAML}, each task integrates a support set along with a query set, both derived from typical deep learning samples. The samples within these sets are randomly sampled from different distributions of users.

The training procedure for the MAML-based framework mainly includes an inner loop and an outer loop. The inner loop updates the parameters of the inner network (denoted as ${\bm \psi}$), while the outer loop optimizes the outer network's parameters (denoted as ${\bm \omega}$). The outer loop is designed to learn an initialization for the inner network. Specifically, before starting the inner loop, the inner network is initialized by the outer network’s parameters. Then, the inner network’s parameters are optimized over $J$ iterations using the Adam optimizer on the support set, guided by the loss function $\mathcal{L}({\bm \psi},{\bm \omega})$.

After completing $J$ iterations, the updated inner network parameters, denoted as ${\bm \psi}_J$, are used in the outer loop. In the outer loop, the parameters of the outer network ${\bm \omega}$ are updated by utilizing the query sets and a meta-loss ($\mathcal{L}_{meta}$), which depends on both ${\bm \psi}_J$ and ${\bm \omega}$. The detailed training procedure is given in $\textbf{Algorithm 2}$.

\begin{algorithm}[h]
	\caption{Training Procedure for MAML-Based Framework}
	\begin{algorithmic}[1]
		\State {\bf Input}:\ batch size of training samples $N_{mb}$, training set with size $B$ denoted as $T_{mt}=\{[S_1,Q_1], [S_2,Q_2],...,[S_B,Q_B]\}$, inner network learning rate $\alpha_1$, outer network learning rate $\alpha_2$.
		\State {\bf Initialize}:\ parameters of the inner network ${\bm \psi}$, parameters of the outer network ${\bm \omega}$.
		\State {\bf for} epoch $=1, 2, ...$ {\bf do}
		\State \quad Sample a batch of samples from $T_{mt}$, and let \Statex \quad \quad $\mathcal{L}_{meta}=0$.
		\State \quad {\bf for} $i=1, 2, ...,N_{mb}$ {\bf do}
		\State \quad \quad {\bf for} $i=1, 2, ...,J$ {\bf do}
		\State \quad \quad \quad ${\bm \psi}_j = {\bm \psi}_{j-1}- \alpha_1 \nabla_{{\bm \psi}_{j-1}}\mathcal{L}({\bm \psi}_{j-1},{\bm \omega};S_i)$;
		\State \quad \quad {\bf end}
		\State \quad \quad $\mathcal{L}_{meta} = \mathcal{L}_{meta} + \mathcal{L}({\bm \psi}_{J},{\bm \omega};Q_i)$;
		\State \quad {\bf end}
		\State \quad ${\bm \omega} = {\bm \omega}- \frac{\alpha_2}{N_{mb}} \nabla_{{\bm \omega}}\mathcal{L}_{meta}$;
		
		\State {\bf end}
		\label{algorithm 2}
	\end{algorithmic}
\end{algorithm}

\begin{figure*}[ht]
	\vspace{3ex}
	\subfloat[Scenario 1: Users distributed only in one area.]{\includegraphics[width=0.5\textwidth]{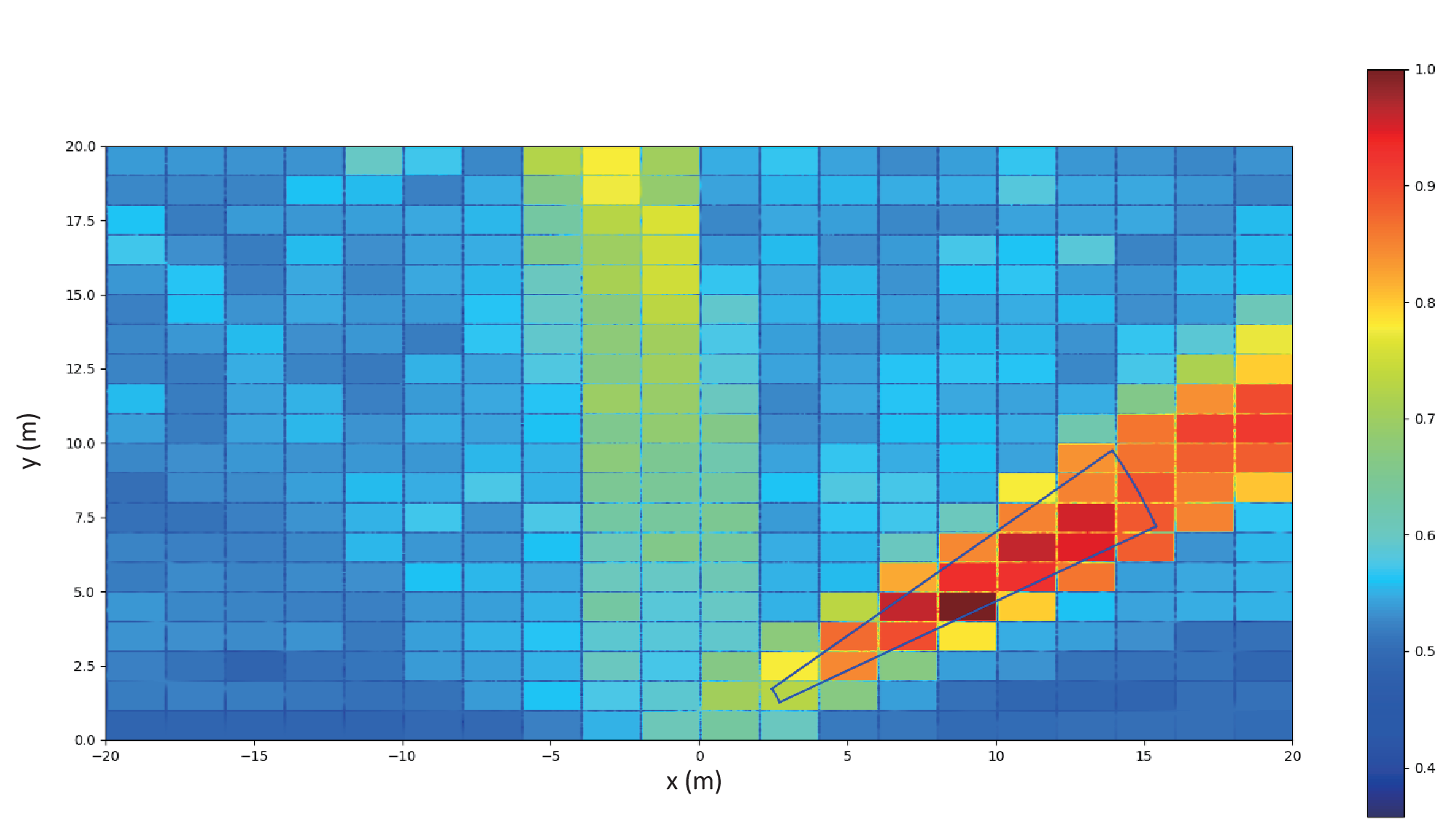}}
	\hfill 	
	\subfloat[Scenario 2: Users distributed in two areas.]{\includegraphics[width=0.5\textwidth]{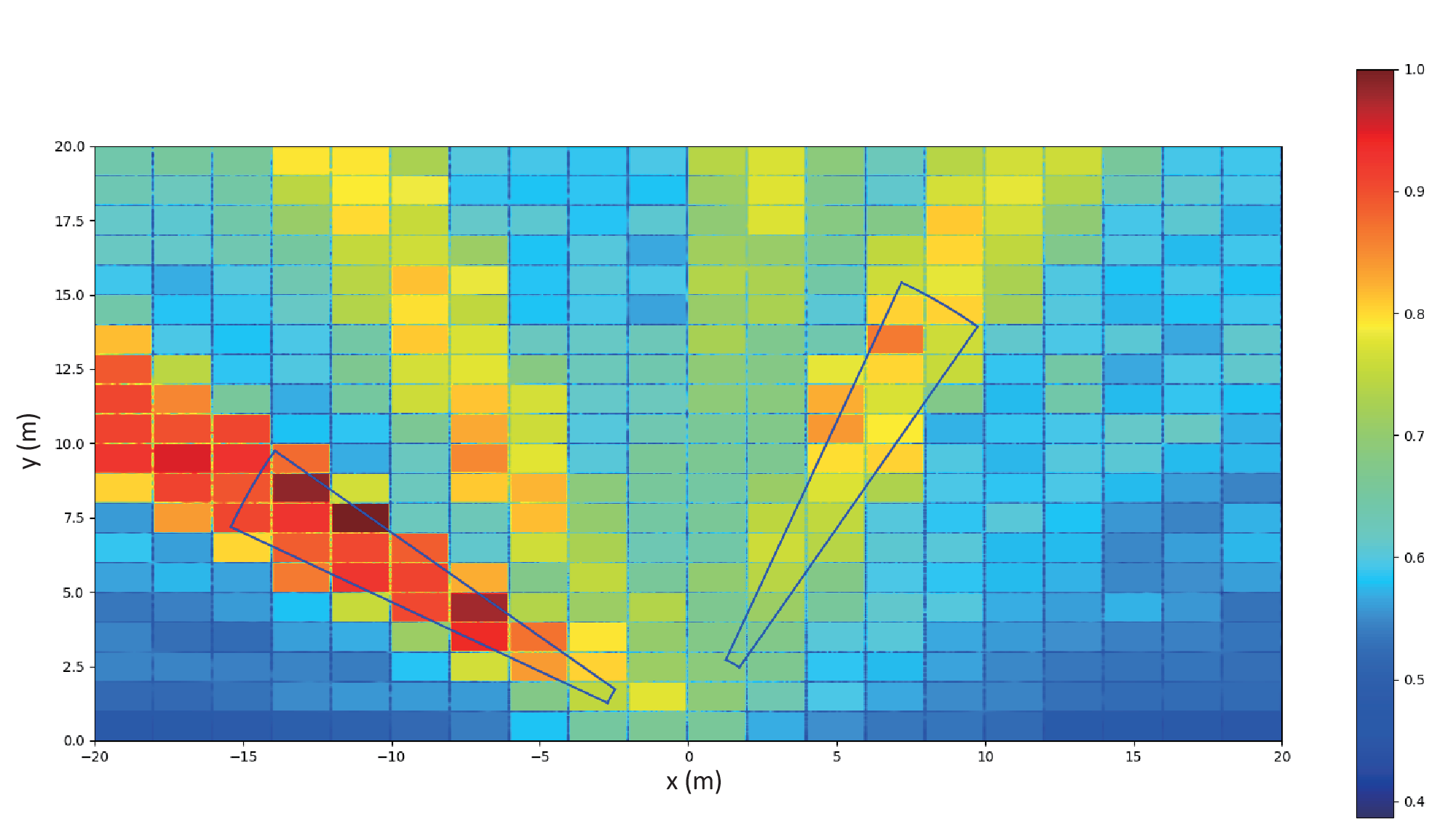}}
	\vspace{3ex}
	\caption{Beam responses for the proposed learned codebook in different scenarios.}
	\label{focus}
	\vspace{0ex}
\end{figure*}

\section{simulation results}
In this section, simulation results are presented to validate the performance of the proposed framework.

\subsection{Simulation Setup and Baseline Schemes}
We assume that the BS is deployed at $(0\text{m},0\text{m},0\text{m})$, and users are randomly located in 2 different areas in the near-field region. In addition, we set $M=4$, $N=1089$, $K=4$, $P_{\text{max}}=30 \text{dBm}$, and $\sigma^2= -10 \text{dBm}$. We compare the proposed meta-learning framework with a traditional codebooks uniformly designed on both the angle and distance and the codebook design DNN introduced in Section III-A.

\begin{figure}[t]
	\vspace{0ex}
	\centering
	\includegraphics[width=1\linewidth]{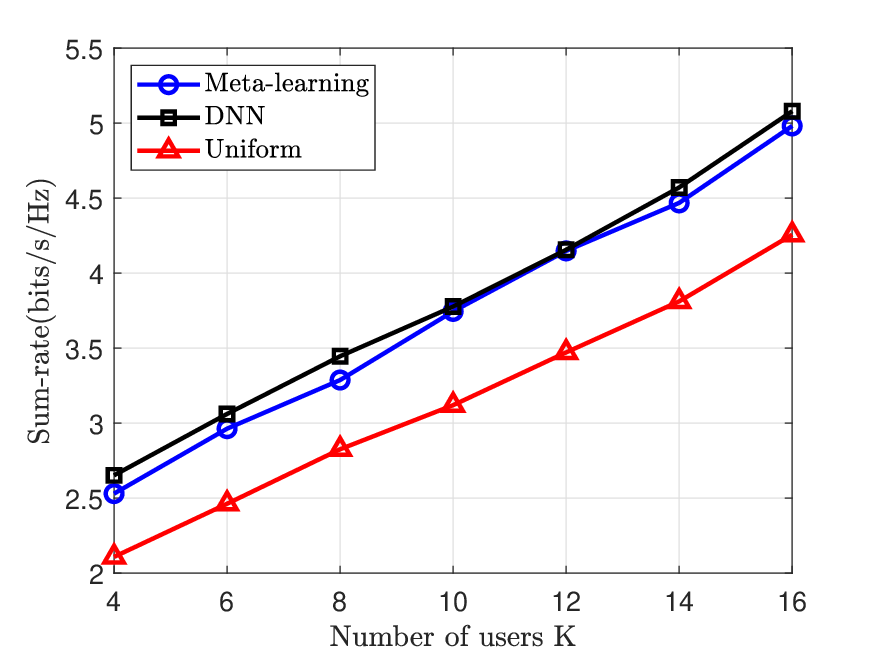}
	\vspace{0ex}
	\caption{Achievable sum-rate versus the number of users $K$.}
	\label{userK}
\end{figure}

\subsection{Performance Evaluation}

To demonstrate the effectiveness of the codebook-learning DNN, we design two distinct test scenarios. In scenario 1, the users are clustered within a single area, while in scenario 2 the users are spread across two different areas. From the results shown in Fig. \ref{focus}, it is evident that the DNN-based codebook aligns well with the user locations. Since the beams in the learned codebook are concentrated on the users' locations rather than covering the entire beam domain, the beam density at the users' locations is increased, leading to improved system performance.

Fig. \ref{userK} depicts the sum-rate versus the number of users $K$. The sum-rate of all approaches increases monotonically with $K$. The proposed framework clearly outperforms the uniform codebook. Additionally, the results show that our framework performs similarly to the DNN, as improved generalization does not always translate into higher performance.

\begin{figure}[t]
	\vspace{0ex}
	\centering
	\includegraphics[width=1\linewidth]{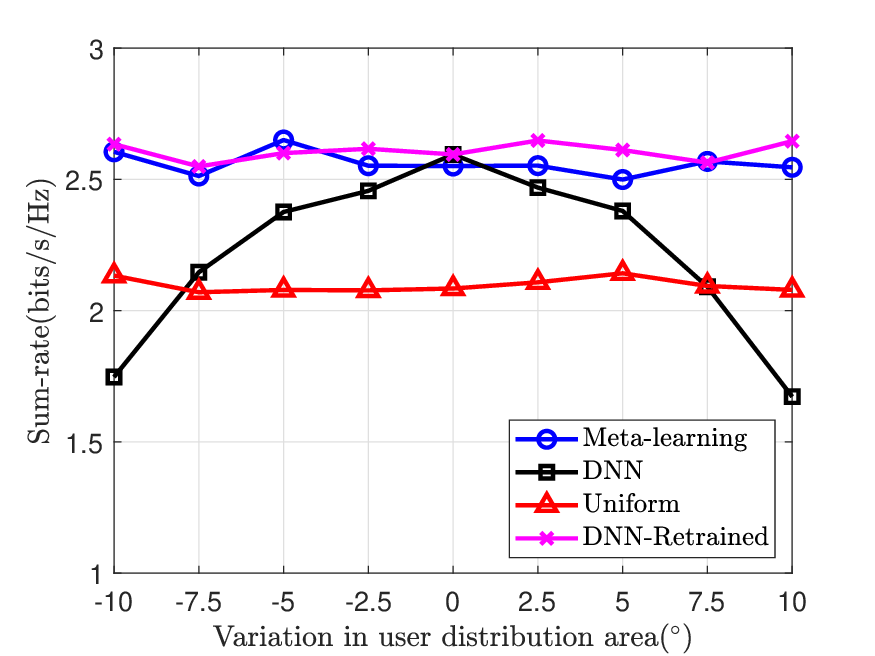}
	\vspace{0ex}
	\caption{Achievable sum-rate versus the variation in user distribution area.}
	\label{degree}
\end{figure}

Fig. \ref{degree} shows the achievable sum-rate as a function of the area over which the users are distributed. The ``DNN-Retrained" approach refers to retraining the DNN for each change in user distribution. The results indicate that the proposed framework exhibits a strong capability for generalization, whereas the DNN struggles with changes in user distribution. Although the DNN-Retrained approach also demonstrates good generalization by adapting to distribution changes, retraining the network for each user distribution is impractical in real-world applications.

\section{conclusion}
In this paper, we investigated a near-field communication system comprising a base station equipped with a UPA and multiple mobile users. Our primary objective was to maximize the system's sum-rate through the optimization of the near-field codebook design. To effectively adapt to user dynamics, we introduced a meta-learning framework that integrates the MAML algorithm with a codebook learning network. 

We initially designed a DNN for learning the near-field codebook. Subsequently, we combined the MAML algorithm with the codebook learning DNN, utilizing the DNN as the inner network. In our proposed framework, the outer network facilitates the inner network in learning effective initializations and enabling rapid adaptation to varying channel conditions. The simulation results demonstrate that our framework significantly outperforms the uniform codebook method and DNN, showcasing superior generalization capabilities.

\vspace{4ex}

\nocite{*}
\bibliographystyle{IEEE}

\begin{thebibliography}{11}
		\vspace{0.5ex}
		
		
\bibitem{NF1}
		Y. Liu, Z. Wang, J. Xu, C. Ouyang, X. Mu, and R. Schober, ``Near-field
		communications: A tutorial review," {\em IEEE Open J. Commun. Soc.}, vol. 4,
		pp. 1999--2049, 2023.

\bibitem{XL1}
Z. Zhang, Y. Xiao, Z. Ma, M. Xiao, Z. Ding, X. Lei, G. K. Karagiannidis,
and P. Fan, ``6G wireless networks: Vision, requirements,
architecture, and key technologies," {\em IEEE Veh. Technol. Mag.}, vol. 14,
no. 3, pp. 28--41, Sep. 2019.

\bibitem{XL2}
E. Bjornson, L. Sanguinetti, H. Wymeersch, J. Hoydis, and T. L.
Marzetta, ``Massive MIMO is a reality—What is next?: Five promising
research directions for antenna arrays," {\em Digital Signal Process.}, vol. 94,
pp. 3--20, Nov. 2019.

\bibitem{XL3}
Y. Liu, X. Liu, X. Mu, T. Hou, J. Xu, M. Di Renzo, and N. Al-Dhahir,
``Reconfigurable intelligent surfaces: Principles and opportunities,"{\em IEEE Commun. Surv. Tut.}, vol. 23, no. 3, pp. 1546--1577, 3rd Quart.
2021.

\bibitem{THz1}
S. Dang, O. Amin, B. Shihada, and M.-S. Alouini, ``What should 6G
be?" {\em Nat. Electron.}, vol. 3, no. 1, pp. 20--29, Jan. 2020.

\bibitem{THz2}
I. F. Akyildiz, J. M. Jornet, and C. Han, ``Terahertz band: Next frontier
for wireless communications," {\em Phys. Commun.}, vol. 12, pp. 16--32, Sep.
2014.
		
\bibitem{NF2}
J. D. Kraus and R. J. Marhefka, {\em Antennas for All Applications.} New York, NY, USA: McGraw-Hill, 2002.

\bibitem{NF3}
K. T. Selvan and R. Janaswamy, ``Fraunhofer and Fresnel distances:
Unified derivation for aperture antennas," {\em IEEE Antennas Propag. Mag.},
vol. 59, no. 4, pp. 12--15, Aug. 2017.

\bibitem{NF4}
D. Slepian and H. O. Pollak, ``Prolate spheroidal wave functions, Fourier
analysis and uncertainty—I," {\em Bell System Techn. Journal}, vol. 40, no. 1,
pp. 43--63, Jan. 1961.

\bibitem{NF5}
D. Slepian, ``On bandwidth," in {\em Proc. of the IEEE}, vol. 64, no. 3, pp.
292--300, Mar. 1976.


\bibitem{NFBF}
H. Zhang, N. Shlezinger, F. Guidi, D. Dardari, and Y. C. Eldar,
``6G wireless communications: From far-field beam steering to near-field beam focusing," {\em IEEE Commun. Mag.}, vol. 61, no. 4, pp. 72--77, Apr. 2023.

\bibitem{NFCE}
M. Cui and L. Dai, ``Channel estimation for extremely large-scale
MIMO: Far-field or near-field?" {\em IEEE Trans. Commun.}, vol. 70, no. 4,
pp. 2663--2677, Apr. 2022.




\bibitem{NFISAC}
D. Galappaththige, S. Zargari, C. Tellambura, and G. Y. Li, ``Near-field ISAC: Beamforming for multi-target detection," {\em IEEE Wireless Commun. Lett.}, vol. 13, no. 7, pp. 1938--1942, Jul. 2024.

\bibitem{NFRIS}
J. Xu, X. Mu, and Y. Liu, ``Exploiting STAR-RISs in near-field communications," {\em IEEE Trans. Wireless Commun.}, vol. 23, no. 3, pp. 2181--2196, Mar. 2024.

\bibitem{HB1}
J. Chen, F. Gao, M. Jian, and W. Yuan, ``Hierarchical codebook design for near-field mmWave MIMO communications systems," {\em IEEE Wireless Commun. Lett.}, vol. 12, no. 11, pp. 1926--1930, Nov. 2023.

\bibitem{HB2}
Y. Zhang, X. Wu, and C. You, ``Fast near-field beam training for
extremely large-scale array," {\em IEEE Wireless Commun. Lett.}, vol. 11, no. 12, pp. 2625--2629, Dec. 2022.
		
\bibitem{DFT}
R. Guo, Y. Cai, M. Zhao, Q. Shi, B. Champagne, and L. Hanzo,
``Joint design of beam selection and precoding matrices for mmWave
MU-MIMO systems relying on lens antenna arrays," {\em IEEE J. Sel. Topics
Signal Process.}, vol. 12, no. 2, pp. 313--325, May 2018.

\bibitem{MLCB0}
M. Alrabeiah, Y. Zhang, and A. Alkhateeb, ``Neural networks based beam codebooks: Learning mmWave massive MIMO beams that adapt to deployment and hardware," {\em IEEE Trans. Commun.}, vol. 70, no. 6,
pp. 3818--3833, Jun. 2022.

\bibitem{MLCB}
F. Liang and Y. Cai, ``Machine learning-enabled joint codebook design and beam selection," in {\em Proc. IEEE Int. Symp. on Wireless Commun. Syst. (ISWCS)}, Rio de Janeiro, Brazil, Jul. 2024, pp. 1--6.

\bibitem{MMS}
A. Sayeed and J. Brady, ``Beamspace MIMO for high-dimensional
multiuser communication at millimeter-wave frequencies," in {\em Proc.
IEEE Global Commun. Conf. (GLOBECOM)}, Atlanta, GA, USA, Dec.
2013, pp. 3679--3684.

\bibitem{ZF}
T. L. Marzetta, ``Noncooperative cellular wireless with unlimited
numbers of base station antennas," {\em IEEE Trans. Wireless Commun.},
vol. 9, no. 11, pp. 3590--3600, Nov. 2010.
		
		
		
	\end{thebibliography}
\begin{footnotesize}

\end{footnotesize}

\end{document}